\documentclass[12pt]{article}
\usepackage{epsfig,fullpage,multirow,amsmath,amsfonts,natbib,latexsym,mathrsfs,enumitem}
\usepackage{bm}

\newtheorem{defn}{Definition}[section]

\begin{document}

\begin{center}
{\Large Bilinear Mixed-Effects Models
 for Affiliation Networks}\\[.3in]
Yanan Jia\footnotemark[1] \hspace{1cm} Catherine A. Calder\footnotemark[1]\footnotemark[3] \hspace*{1cm} Christopher R. Browning \footnotemark[2] \hspace{2cm} \\[.2in]
\today
\end{center}

\footnotetext[1]{Department of Statistics, The Ohio State University, Columbus, OH, USA} 
\footnotetext[2]{Department of Sociology, The Ohio State University, Columbus, OH, USA} 
\footnotetext[3]{Email:  calder@stat.osu.edu}

\begin{abstract}   
An affiliation network is a particular type of two-mode social network that consists of a set of `actors' and a set of `events' where ties indicate an actor's participation in an event.  Although networks describe a variety of consequential social structures, statistical methods for studying affiliation networks are less well developed than methods for studying one-mode, or actor-actor, networks.  One way to analyze affiliation networks is to consider one-mode network matrices that are derived from an affiliation network, but this approach may lead to the loss of important structural features of the data.  The most comprehensive approach is to study both actors and events simultaneously. In this paper, we extend the bilinear mixed-effects model, a type of latent space model developed for one-mode networks, to the affiliation network setting by considering the dependence patterns in the interactions between actors and events and describe a Markov chain Monte Carlo algorithm for Bayesian inference.  We use our model to explore patterns in extracurricular activity membership of students in a racially-diverse high school in a Midwestern metropolitan area.  Using techniques from spatial point pattern analysis, we show how our model can provide insight into patterns of racial segregation in the voluntary extracurricular activity participation profiles of adolescents.   

{\bf Keywords:} Bayesian modeling, generalized linear model, social networks, Markov chain Monte Carlo (MCMC), latent space, point pattern, racial segregation, visualization

\end{abstract}
\section{Introduction}
\label{se:data}

In typical statistical analyses, the primary goal is learning about properties of individual units.
When the property of interest involves interactions between multiple units rather than properties of the individual units themselves, the units can be considered a network. Network data are widely used to represent relational information among interacting units.  Units are referred to as \textit{nodes} in a network, and relationships between the nodes are represented by \textit{ties/edges}.  Pairs of nodes, which may be either linked or not, are called \textit{dyads} in a network.  

We use the term  \textit{mode}  to differentiate sets of distinct nodes in a network.
The most common type of network is a one-mode network in which all units are of the same type. A typical example is a friendship network where all nodes are individuals, or \textit{actors}, and ties between all actors are well defined.   Two-mode networks contain relational information about two distinct sets of entities, specifically about ties between nodes of different modes. Two-mode networks can capture more relational structure than the standard one-mode representation of such data and are a natural representation of relational data involving affiliations between sets of entities.  The term ``affiliation'' usually refers to membership or participation data. Arguably, the most well known affiliation dataset is the ``Southern Women" network collected by  \cite{davis}, which consists of attendance records at various social events in a small southern town. This dataset is an affiliation network since the ties represent affiliations between a set of \textit{actors} (women), denoted by $A$, with a set of \textit{events} (social events), denoted by $E$.   Affiliation networks, such as the Southern Women network, allow the study of the dual perspectives of actors and events where connections among members of one of the modes are based on linkages established through the second mode (i.e., women are connected because they attend the same social events and social events are connected through the women that participate in them).  In this paper, we study patterns of participation in extracurricular activities within a racially diverse high school in a Midwestern metropolitan area of the United States.  In particular, we aim to identify patterns of racial segregation in the extracurricular activity profiles of students.   Details on the data and aims of our segregation analysis are provided in Section \ref{se:description}.

In this paper, we build  on ideas from \cite{hoff} and extend the bilinear mixed-effects models developed for one-mode networks to the two-mode settings. The bilinear effect for an actor/event pair in our model is the inner product of unobserved characteristic vectors specific to actors and events.  Our model can capture fourth order (even number order) dependence, which we argue below is necessary to describe the types of structures seen in real affiliation networks.  Inferences from our model provide a visual and interpretable model-based spatial representation of affiliation relationships.  If we presume the existence of a latent \emph{social space} in which the positioning of actors captures similar profiles of event participation, these latent positions allow us to explore, as well to test hypotheses about, social structure within an affiliation network.  Here, we describe how methods from spatial point pattern analysis can be used to investigate the presence of racial segregation in extracurricular activity memberships of high school students.  

This paper is organized as follows. In the next section, we introduce types of dependence often seen in two-mode network datasets, discuss basic models for affiliation network data, and argue that these basic models are not sufficiently able to capture the dependencies in two-mode network data. In Section \ref{se:models},  we state our bilinear mixed effects model and demonstrate how it is able to capture higher-order dependence than standard mixed-effects models.  We also provide a description of a Markov chain Monte Carlo (MCMC) algorithm providing full Bayesian inference. Our analysis of student extracurricular activity participation is presented in Section \ref{se:application}.  We conclude in Section \ref{se:discussion} with a discussion of some directions for future research.

\section{Affiliation Networks}\label{se:affiliation_networks}

\subsection{Background}\label{se:background}
Generally, an affiliation network can be denoted by an $n^a \times n^e$  affiliation matrix $\mathbf{Y}=\{ y_{ik} \}$, which records the affiliation of each actor with each event, where rows index actors and columns index events, and $n^a$ and $n^e$ are the total number of actors and events, respectively. The entries of this matrix, $y_{ik}$, can be binary variables or non-negative integer-valued variables. If actor $i$ is affiliated with event $k$, then $y_{ik} \geq  1$  and $y_{ik}=0$ otherwise, where $i=1,2,\dots, n^a$ and $k=1,2,\dots, n^e$. Each row of  $\mathbf{Y}$ describes an actor's affiliation with the events. Similarly, each column of  $\mathbf{Y}$ describes the membership of an event.
An affiliation network can also be represented by a  bipartite graph, or a graph in which  the nodes can be partitioned into two subsets corresponding to the distinct modes, and all lines are between pairs of nodes belonging to the different modes. For affiliation networks, since actors are affiliated with events, and events have actors as members, all lines in the bipartite graph are between nodes representing actors and nodes representing events.  

Statistical methods for one-mode networks are fairly well developed. The exponential random graph model (ERGM) is one of the most popular methods for analyzing networks  \citep{frank,wasserman, pattison,robins}.
Although ERGMs are useful for modeling global network characteristics, they are known to possess some undesirable properties. 
\cite{robins} and \cite{Handcock03assessingdegeneracy} discussed  these challenges associated with ERGMs, including the intractability of the normalizing  constant in the likelihood function of ERGMs and model degeneracy.  \citet{snijders:etal:2006} proposes an alternative specification of ERGMs that partially addresses these issues, but requires specifying values of tuning parameters.  As an alternative, models built on latent variables have attracted considerable attention recently.  These models include mixed-effects models  \citep{van, zijlstra,hoff3,hoff,hoff9}, 
the stochastic blockmodel  \citep{wangwong, snijders7, snijders1}, and latent space models \citep{hoff2}. All of these latent variable models  assume conditional independence of the probability of ties between dyads. That is, the elements of $\mathbf{Y}$ are independent conditional on latent variables.  Conditional independence does not imply that latent variable models cannot capture network dependencies of interest. Indeed, some of the more sophisticated latent variable models make clever use of latent structures to capture types of dependence. The conditional independence of edges implies that model degeneracy is not an issue.  In addition, the conditional independence of tie probabilities leads to  computational advantages in model fitting \citep{computer}.
A latent variable model that we build on in this paper is the bilinear mixed-effects model proposed by \cite{hoff}, which is an extension of latent space models for one-mode networks. This model uses interacting latent variables to capture certain types of higher-order dependence patterns often present in social networks.

While there is a rich literature on statistical methods for one-mode networks, methods for two-mode networks are limited. One approach, known as the ``conversion," or projection method \citep{proj}, relies on the two one-mode networks that can be derived from an affiliation network: $\mathbf{YY}^{\prime}$ is the one-mode network for actors and $\mathbf{Y}^{\prime}\mathbf{Y}$ is the one-mode network for events. Information is lost, however, by converting an affiliation network into two one-mode networks. For instance, if we use binary matrices to represent the one-mode networks, then we lose information about both the number and the properties of the shared partners of the other set. 
Alternatively, we can build models for two-mode networks to analyze both actors and events simultaneously. \cite{wang} extended ERGMs to the two-mode situation. However, these models suffer  from the limitation of the one-mode ERGMs described above.  In addition, they do not readily permit the modeler to investigate patterns in activity participation across multiple events (e.g., whether certain individuals share activity profiles).  As we will illustrate, the latent variable approach we take is much more amenable to this sort of study.

\subsection{Dependence Patterns in Affiliation Networks}
\label{se:dp} 
Network data differ from other types of dependent data in that ties often tend to be \textit{ transitive, balanced}, and \textit{clusterable} \citep{WF}. In one-mode friendship networks, we often see patterns that indicate ``a friend of a friend is a friend," a statement that translates to properties of sets of three dyads (triangles).  In particular, this pattern is called transitivity.  Balance is a generalized version of transitivity defined for signed relationship of the type $A_{ij}$ is positive if there is a tie between nodes $i$ and $j$ and is negative otherwise.  
Formally, a signed relationship between nodes is denoted as follows:
\[A_{ij} = \left\{ 
  \begin{array}{r l}
    1 & \quad \text{if $i$  and  $j$ are tied, }\\    
    -1 & \quad \text{if $i$ and  $j$ are not tied.}
  \end{array} \right.\]
In one-mode networks, a triangle formed by a triad of units $i, j, k$ is said to be balanced if $A_{ij} \times A_{jk}\times A_{ki} > 0$. 
Clusterability is a generalization of the concept of balance. A triangle is clusterable if it is either balanced or the pairwise relationships within the triad are all negative.  

Here we extend these definitions to the two-mode setting, which to the best of our knowledge has not been done previously.  We say a set of four possible ties among a \emph{tetrad} of units consisting of one pair of actors $i,j$ and one pair of events $k,l$, $  \{A_{ik} , A_{il} , A_{jk} ,A_{jl} \}$, forms a \emph{cycle}, and offer the following definitions.
\begin{defn}
For signed affiliation relations, a cycle $  \{A_{ik} , A_{il} , A_{jk} ,A_{jl} \}$  is transitive if whenever $A_{ik}=A_{il}=A_{jk}=1$, we have $A_{jl} =1$. 
\end{defn} 
Transitivity implies that if  actors $i$ and $j$ both have a tie with event $k$ and actor $i$ is tied with another event $l$, then we expect actor $j$ also has a tie with event $l$. 
\begin{defn}
For signed affiliation relations,  a cycle $  \{A_{ik} , A_{il} , A_{jk} ,A_{jl} \}$ is said to be balanced if $A_{ik} \times A_{il} \times A_{jk} \times A_{jl} =1$. 
\end{defn}
Since the number of elements in a cycle is an even number (4), we note that balance and clusterability are identical conceptually in the two-mode setting. 

\begin{figure}[htp]\begin{center} 
\includegraphics[width=6in]{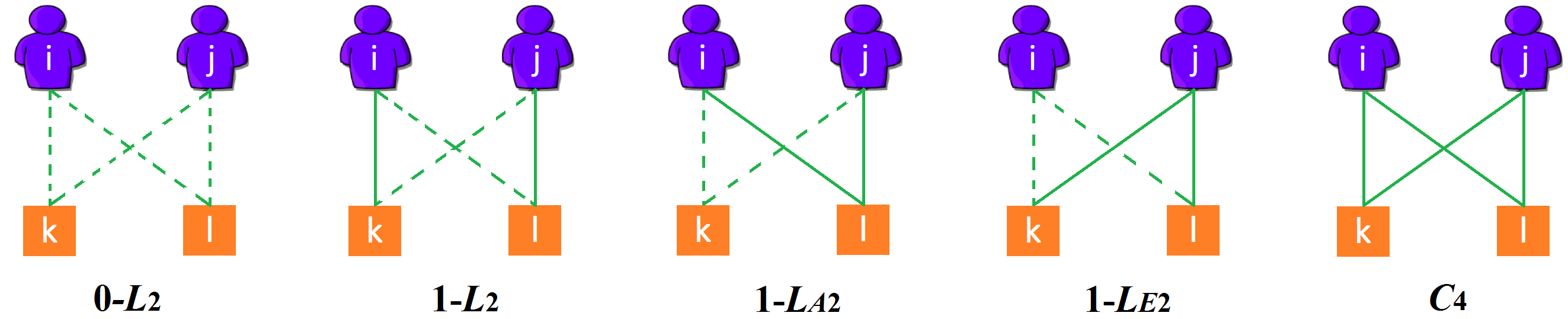}
\caption{All possible balanced cycles among a tetrad.  Solid lines connecting actors and events denote ties (positive relationship) and dashed lines denote the absence of ties (negative relationship).}
\label{fig:balance}
\end{center}\end{figure}

For general signed relations among units, many theories of social systems suggest that the relationships within a cycle tend to be balanced.  For example, if $A_{ik}=1$ and $A_{jk}=1$, which means the relationships between actor $ i$ and event $k$ and between actor $j$ and event $k$ are positive, then it is more likely that either both $A_{il}=1$ and $A_{jl}=1$ or both $A_{il}=-1$ and $A_{jl}=-1$.  In other words, if actors $i$ and $j$ both participate in event $k$, then they are likely to either both participate in event $l$ or both not participate in event $l$. In real affiliation networks, we expect to see more evidence of balance than we would expect if the presence of ties is completely random.  This translates into the presence of particular balanced patterns among cycles, which are illustrated in Figure \ref{fig:balance}.  

These configurations, $0$-two-path $(0$-$ L_{2})$, $1$-two-path $(1$-$L_{2})$, actor $1$-two-path $(1$-$L_{A2})$, event $1$-two-path $(1$-$L_{E2})$, and four-cycles $(C_4)$, shown in Figure \ref{fig:balance} are the balanced cycles often seen in affiliation networks.  Generalizations of these structures, the actor $k$-two-path $(k$-$L_{A2})$  and the event $k$-two-path $(k$-$L_{E2})$, are used by  \cite{wang} to define ERGMs for two-mode networks.

Consider the case where actor-event ties within an affiliation network are assumed to be independent and identically distributed with tie probability $\pi_0$.  In this case, it can be shown that the expected proportion of balanced cycles is $\pi =  \pi_0^4+(1-\pi_0)^4+6\pi_0^2(1-\pi_0)^2$.  In many real networks, the observed proportion of balanced cycles is greater than this theoretical value from this model (i.e., independent and identically distributed ties).  That is, $p >  p_0^4+(1-p_0)^4+6p_0^2(1-p_0)^2$, where $p$ is the observed proportion of balanced cycles and $\pi_0$ is the observed proportion of actor-event pairs that are tied.  From this, we can see the importance of capturing fourth-order dependence (dependence between tetrads) in models for affiliation network data.

\subsection{Basic Models}\label{se:bm} 

Our data consist of an $n^a \times n^e$ sociomatrix $\mathbf{Y}$, with entries $y_{ik}$ denoting the value of the relation between actor $i$ and event $k$ and  additional covariate information associated with actors, events, and dyads.   
 
\subsubsection{Fixed-Effects Model}
Since most affiliation network data, $y_{ik}$, are binary or (non-negative) integer valued,  we specify models using the standard generalized linear model framework.  We let
\[Pr(\mathbf{Y} = \mathbf{y} \vert  \pmb{\beta}) = \prod_{i = 1}^{n^a} \prod_{k=1}^{n^e}Pr(Y_{ik} = y_{ik} \vert \pmb{\beta}), \]
where each component of $\mathbf{Y}$ follows an exponential family distribution.  We relate $\mu_{ik} \equiv E(Y_{ik} \vert \pmb{\beta})$ to a set of covariate variables $\mathbf{x}_{ik}$ via a link function denoted by $g(\cdot)$:
\[
\theta_{ik} =g(\mu_{ik})=\pmb{\beta}^{\prime} \mathbf{x}_{ik}, 
\]
where $\pmb{\beta}$ is a $r$-dimensional vector of unknown regression coefficients. 
We decompose $\mathbf{x}_{ik}$ into  $\mathbf{x}_{ik}= ( \mathbf{x}_{ik}^d, \mathbf{x}_{i}^a, \mathbf{x}_{k}^e )$, where $\mathbf{x}_{ik}^d$ is an $r^d$-dimensional covariate vector associated with (actor $i$, event $k$) dyad, 
$\mathbf{x}_{i}^a$ is an  $r^a$-dimensional covariate vector associated with  actor $i$, $\mathbf{x}_{k}^e $ is an  $r^e$-dimensional covariate vector associated with event $k$, implying $r^d+r^a+r^e=r$. The model can then be rewritten as
\begin{equation}
\theta_{ik}=g(\mu_{ik})=\pmb{\beta}_d^{\prime} \mathbf{x}_{ik}^d +\pmb{\beta}_a^{\prime} \mathbf{x}_{i}^a +\pmb{\beta}_e^{\prime} \mathbf{x}_{k}^e.
\end{equation}
where $\pmb{\beta}_d, \pmb{\beta}_a$, and $\pmb{\beta}_e$ are vectors of unknown regression coefficients with dimension $r^d, r^a$ and $r^e$, respectively.
The affiliation network data are measured on a set of actors and a set of events. Since actors and events comprise multiple dyads, the observations  $y_{ik}$s are likely not (conditionally) independent given the regression coefficients, and we need a model which can capture dependence induced by the shared actors and events making up the dyads.

\subsubsection{Mixed-Effects Models}\label{se:mixedmodels}
For affiliation network data, an actor can attend multiple events and an event can have multiple actors.  To model the within-node dependence, we consider mixed models with actor and event random effects of the form 
\begin{equation}
\theta_{ik}=g(\mu_{ik})=\pmb{\beta}_d^{\prime} \mathbf{x}_{ik}^d +\pmb{\beta}_a^{\prime} \mathbf{x}_{i}^a +\pmb{\beta}_e^{\prime} \mathbf{x}_{k}^e+ a_i+e_k,
\label{eq:nogamma}
\end{equation} 
where $\mu_{ik} \equiv E(Y_{ik} \vert \theta_{ik} )$, and $a_i$ and $e_k$ represent the actor and event random effects, respectively. 
For discrete data subject to overdispersion (Poisson, binomial), an observation level residual is also present, so that,
\begin{equation}
\theta_{ik}=g(\mu_{ik})=\pmb{\beta}_d^{\prime} \mathbf{x}_{ik}^d +\pmb{\beta}_a^{\prime} \mathbf{x}_{i}^a +\pmb{\beta}_e^{\prime} \mathbf{x}_{k}^e+ a_i+e_k+ \gamma_{ik},
\label{eq:gamma}
\end{equation}
with $\gamma_{ik}$ usually taken as independent and identically distributed errors \citep{abhm}.  We can interpret the $\gamma_{ik}$s as dyad random effects. 

The observations $\{Y_{ik}: i=1, \dots, n^a, k=1, \dots, n^e \}$ are modeled as conditionally independent given the random effects, denoted by $\mathbf{a} = (a_1, \dots, a_{n^a})^{\prime}$, $\mathbf{e} = (e_1, \dots, e_{n^e})^{\prime}$, and $\pmb{\gamma} = \text{vec}(\mathbf{\Gamma})$ for the $n^a \times n^e$ matrix $\mathbf{\Gamma}$ with elements $\gamma_{ik}$ for $i = 1, \dots, n^a$ and $k=1, \dots, n^e$. That is,
\[Pr(\mathbf{Y} = \mathbf{y} \vert  \pmb{\beta}, \mathbf{a}, \mathbf{e}, \pmb{\gamma}) = \prod_{i = 1}^{n^a} \prod_{k=1}^{n^e}Pr(Y_{ik} = y_{ik} \vert \pmb{\beta}, a_i, e_k, \gamma_{ik}). \]
We take the different types of random effects to be mutually independent and Gaussian with mean zero and variances $\sigma^2_a$, $\sigma^2_e $ and $\sigma^2_{\gamma}$, respectively:
\[\mathbf{a} \, \vert \, \sigma^2_a \sim \text{MVN} ( 0, \sigma^2_a \mathbf{I}_{n^a \times n^a} ),\] 
\[\mathbf{e}  \, \vert \, \sigma^2_e \sim \text{MVN} ( 0, \sigma^2_e \mathbf{I}_{n^e \times n^e} ),\]
and
\[\pmb{\gamma}  \, \vert \, \sigma^2_{\gamma} \sim \text{MVN} ( 0, \sigma^2_{\gamma} \mathbf{I}_{n^{\gamma} \times n^{\gamma}} ),\]
where $\mathbf{I}_{n \times n}$ denotes the $n$-dimensional identity matrix and  $n^{\gamma}=n^a \times n^e$. The model given by (\ref{eq:nogamma}) is a special case of (\ref{eq:gamma}), where $\sigma_{\gamma}^2$ is equal to zero.  Therefore, we refer to the model given by (\ref{eq:gamma}) as the generalized linear mixed effects model for affiliation networks.

Letting $\epsilon_{ik}$ denote the $(i,k)$ random effect  (i.e., $\epsilon_{ik} = a_i+e_k+\gamma_{ik}$) and marginalizing over the $a_i$s, $e_k$s, and $\gamma_{ik}$s, it follows that 
\begin{align*}
\mbox{Cov}(y_{ik}, y_{il}  ) &=\mbox{E}( \epsilon_{ik}  \epsilon_{il} )= \sigma_a^2  \\
\mbox{Cov}(y_{ik}, y_{jk} ) &=\mbox{E}( \epsilon_{ik}  \epsilon_{jk} )= \sigma_e^2   \\
\mbox{Cov}(y_{ik}, y_{ik} ) &=\mbox{E}( \epsilon_{ik} ^2 )= \sigma_a^2+ \sigma_e^2 + \sigma_ {\gamma}^2 
\end{align*}
where $\sigma_a^2$ and $\sigma_e^2$ capture the components of the total variation in the $\epsilon_{ik}$s explained by dyads containing the same actor or event, respectively.  This model is able to capture dependence between elements of $\mathbf{Y}$ due to shared nodes using a standard random-effects specification. However, as we will discuss in the next section, this mixed model is unable to capture the fourth order (or higher even order) dependence frequently encountered in real affiliation networks.  

\section{A Bilinear Mixed-Effects Model}
\label{se:models} 

\subsection{Model Specification}\label{specification}
In order to capture more transitivity and balance than the generalized linear mixed effects model allows, we add a bilinear random effect to the model given by (\ref{eq:gamma}).  As with \citet{hoff}'s bilinear mixed effects model for one-mode network data, this addition enables us to capture the expected balanced tendencies in two-mode network relations. 

We presume the existence of a latent social space of dimension $t$.  Both actors and events have positions in this latent space, denoted by the vectors $\mathbf{u}_i$ and $\mathbf{v}_k$, respectively.   If we consider the pair of actors with position vectors $(\mathbf{u}_i, \mathbf{u}_j)$ (or the pair of events with position vectors $(\mathbf{v}_k, \mathbf{v}_l)$), and they have similar direction and magnitude, then the inner products $\mathbf{u}_i ^{\prime} \mathbf{v}_k $  and $\mathbf{u}_j^{ \prime} \mathbf{v}_k $   (or  $\mathbf{u}_i ^{\prime} \mathbf{v}_k $  and $\mathbf{u}_i ^{ \prime} \mathbf{v}_l $ ) will not be too different. 
A probability measure over these unobserved characteristics induces a model in which the presence of a tie between an actor and an event is dependent on the presence of other ties. Relations modeled as such are probabilistically balanced. 

We add this inner product of latent vectors $ \mathbf{u}_i$ and $\mathbf{v}_k $ to (\ref{eq:gamma}) so that
\[
\epsilon_{ik}= a_i+e_k  + \gamma_{ik}+ \varepsilon_{ik},
\]
where the random effects $a_i, e_k$ and $\gamma_{ik}$ are still taken to be multivariate normal with means and covariances are as given in Section \ref{se:mixedmodels}. This set of bilinear terms
 $\{\varepsilon_{ik}= \mathbf{u}_i ^{ \prime} \mathbf{v}_k, i= 1, \dots, n^a, k = 1, \dots, n^e \}$ 
  allows us to capture balance.  To see this further, consider the case in which $t=1$, where $t$ is the dimension of the $\mathbf{u}_i$ and $\mathbf{v}_k$ vectors.  In this case, the $\varepsilon_{ik}$s correspond to the residuals from the version of the model without the bilinear term.  Since $\varepsilon_{ik} \times \varepsilon_{il} \times \varepsilon_{jk} \times \varepsilon_{jl} = (u_i u_j v_k v_l)^2 \geq 0$, the bilinear term can be seen to capture positive residual cycles.  Of course in a real dataset, we do not expect networks to be completely balanced.  By taking $t >1$, the bilinear term captures the balanced tendencies of real networks without forcing every residual cycle to be positive. 
  
We assume the  $\mathbf{u}_i$s and $\mathbf{v}_k$s are mutually independent and follow $t$-dimensional multivariate normal distributions so that  
\[
\mathbf{u}_i \, \vert \, \pmb{\Sigma}_u \stackrel{}{\sim} \text{MVN} (0, \pmb{\Sigma}_u ) \hspace{20mm}  
\mathbf{v}_k \,\vert \, \pmb{\Sigma}_v \stackrel{}{\sim}\text{MVN} (0, \pmb{\Sigma}_v ).
\]
In addition, we assume $\mathbf{u}_i \perp \mathbf{u}_j$ for $\{i,j = 1, \dots, n^a: i \neq j\}$ and $\mathbf{v}_k \perp \mathbf{v}_l$ for $\{k,l = 1, \dots, n^e: k \neq l\}$.
It follows that $\varepsilon_{ik}$s have moments
\[
\mbox{E}(\varepsilon_{ik})=0
\]
\[ 
\mbox{E}(\varepsilon_{ik}^2)= \text{trace} (\pmb{\Sigma}_u \pmb{\Sigma}_v)
\]
\[
\mbox{E}(\varepsilon_{ik}  \varepsilon_{jk} \varepsilon_{jl} \varepsilon_{il})=\text{trace} (\pmb{\Sigma}_u  \pmb{\Sigma}_v \pmb{\Sigma}_u  \pmb{\Sigma}_v).
\]
The other second, third, and fourth order moments are all equal to zero.
For simplicity, we assume 
$\pmb{\Sigma}_u=\sigma_u^2 \mathbf{I}_{t \times t},\hspace{2mm} 
 \pmb{\Sigma}_v={\sigma}_v^2 \mathbf{I}_{t \times t}$.
In this case, the moments of bilinear term become
\begin{align*}
\hspace{10mm} \mbox{E}(\varepsilon_{ik}^2)=t  {\sigma}_u^2  {\sigma}_v^2, 
\hspace{20mm}  
\mbox{E}(\varepsilon_{ik}  \varepsilon_{jk} \varepsilon_{jl} \varepsilon_{il})=t \sigma _u^4  {\sigma}_v^4. 
\end{align*}
This gives the following nonzero second and forth order moments for the bilinear random-effects components, $
\epsilon_{ik}= a_i+e_k  + \gamma_{ik}+ \mathbf{u}_i ^{ \prime} \mathbf{v}_k$:
\begin{align*}
\mbox{E}( \epsilon_{ik}  \epsilon_{il} )=  \sigma _a^2,   
& \hspace{20mm}
\mbox{E}( \epsilon_{ik}  \epsilon_{jk} )=  {\sigma}_e^2,  \\ 
\mbox{E}( \epsilon_{ik} ^2 )=  {\sigma}_a^2+ \ {\sigma}_e^2  +  {\sigma}_{\gamma}^2+ t {\sigma}_u^2  {\sigma}_v^2,
& \hspace{20mm}
\mbox{E}(\epsilon_{ik}  \epsilon_{jk} \epsilon_{jl} \epsilon_{il})= {\sigma}_a^4+ {\sigma}_e^4  +t {\sigma}_u^4  {\sigma}_v^4.
\end{align*}
The bilinear effect $\varepsilon_{ik}=\mathbf{u}_i ^{\prime} \mathbf{v}_k $ can be interpreted as a mean-zero random effect that is able to capture particular fourth order dependence in affiliation network data.

\subsection{Parameter Estimation}
\label{se:parameter}
 
The parameters we want to estimate are $\{ \pmb{\beta}_d, \pmb{\beta}_a, \pmb{\beta}_e, {\sigma}_a^2,  {\sigma}_e^2, {\sigma}_{\gamma}^2,  {\sigma}_u^2,  {\sigma}_v^2 \}$.  Following \cite{hoff}, we work with the following representation of our model:
\begin{align}
\theta_{ik} &= \pmb{\beta}_d ^{\prime} \mathbf{x}^d_{ik} + ( \pmb{\beta}_a^{ \prime} \mathbf{x}^a_{i}+a_i )+ ( \pmb{\beta}_e^{ \prime} \mathbf{x}^e_{k}+e_k ) + \gamma_{ik}+ \mathbf{u}_i ^{\prime} \mathbf{v}_k \nonumber\\
&= \pmb{\beta}_d ^{\prime} \mathbf{x}^d_{ik} +{\mu^a_i} + {\mu^e_k}  + \gamma_{ik}+ \mathbf{u}_i ^{\prime} \mathbf{v}_k, 
\label{eq:bilinear}
\end{align}
where
$
\mu^a_i = \pmb{\beta}_a ^{\prime} \mathbf{x}_{i}^a+a_i 
$
and 
$
\mu^e_k = \pmb{\beta}_e ^{\prime} \mathbf{x}_{k}^e+e_k 
$
can be viewed as actor and event specific effects, respectively.  We then define
$
 z_{ik}=\theta_{ik} -\mathbf{u}_i ^{\prime} \mathbf{v}_k = \pmb{\beta}_d ^{\prime} \mathbf{x}^d_{ik} +  \mu^a_i+\mu^e_k  + \gamma_{ik},  
$
and let $\mathbf{z} = \text{vec}(\mathbf{Z})$, where $\mathbf{Z}$ is the $n^a \times n^e$ matrix with elements $z_{ik}$ for $i = 1, \dots, n^a$ and $k=1, \dots, n^e$.  We take $\pmb{\theta}$ to be the $n^a \times n^e$ matrix with elements $\theta_{ik}$, and let $\mathbf{u}$ be a $n^a \times t$ matrix with rows $\mathbf{u}_i$ for $i = 1, \dots, n^a$ and $\mathbf{v}$ be a $n^e \times t$ matrix with rows $\mathbf{v}_i$ for $i = 1, \dots, n^e$.  Then we can write
\begin{equation}\label{eq:bilinear2}
\mathbf{z} =  \text{vec} (\pmb{\theta} -\mathbf{u} \mathbf{v}^{\prime} )=
\mathbf{X}_D
\left( 
\begin{array}{c}
{\pmb{\beta}}_d \\
\pmb{\mu}^a\\
\pmb{\mu}^e 
\end{array} 
\right)
 +
\pmb{\gamma}
\end{equation}
where $\mathbf{X}_D$ is the appropriate design matrix constructed using (\ref{eq:bilinear}) and $\pmb{\gamma}$ is a vector with dimension $n^{\gamma}$ as described in Section \ref{se:mixedmodels}.  From (\ref{eq:bilinear2}), it is clear that conditional on the $\pmb{\theta}$s, $\mathbf{u}$s and $\mathbf{v}$s, the other parameters  can be sampled using a standard Bayesian normal-theory regression approach.

Our general Gibbs sampler given below is similar to \cite{hoff}'s algorithm, with the exception of the first part of step 1 and all of steps 2 and 3 in the outline below:

\begin{enumerate}
\item { Sample linear effects:}\\
Sample $\pmb{\beta}_d, \pmb{\mu}^a, \pmb{\mu}^e \vert \pmb{\beta}_a, \pmb{\beta}_e,  {\sigma}^2_a,  {\sigma}^2_e , {\sigma}_{\gamma}^2, \pmb{\theta}, \bf{u}, \bf{v} $ (linear regression) \\
Sample $\pmb{\beta}_a, \pmb{\beta}_e \vert \pmb{\mu}^a, \pmb{\mu}^e,  {\sigma}^2_a,  {\sigma}^2_e, {\sigma}_{\gamma}^2 $ (linear regression) \\
Sample $ {\sigma}_a^2, {\sigma}_e^2 $, and $ {\sigma}_{\gamma}^2$ from their full conditionals 
\item  {Sample bilinear effects:}\\
For $i=1, \dots ,n_a $ sample $\mathbf{u}_i \vert \mathbf{u}_{-i}, \mathbf{v}, \pmb{\theta}, \pmb{\beta}_d, \pmb{\mu}^a, \pmb{\mu}^e,  {\sigma}_u^2 $  (linear regression ) \\
For $k=1, \dots ,n_e $ sample $\mathbf{v}_k \vert \mathbf{v}_{-k}, \mathbf{u}, \pmb{\theta}, \pmb{\beta}_d, \pmb{\mu}^a, \pmb{\mu}^e ,  {\sigma}_v^2  $  (linear regression ) \\
Sample $ {\sigma}_u^2$  and  ${\sigma}_v^2$ from their full conditionals
\item {Update $\pmb{\theta}  $}: For actor $i$ and event $k$ \\
Propose $\theta_{ik}^*  \sim $ \text{N}$( \bm{\beta}^{\prime} X_{ik} + \mu^a_i +\mu^e_k + \mathbf{u}_i^{\prime}\mathbf{v}_k,  {\sigma}_{\gamma}^2)$\\
Accept $\theta_{ik}^*$ with probability $\left[ {p(y_{ik} \vert \theta_{ik}^*)}/{p(y_{ij} \vert \theta_{ik})} \right] \bigwedge 1$
\end{enumerate}
The full conditional distributions of $ \pmb{\beta}_d, \pmb{\mu}^a,\pmb{\mu}^e $, $ \pmb{\beta}_a, \pmb{\beta}_e $, $ {\sigma}^2_a,  {\sigma}^2_e,  {\sigma}^2_{\gamma}, {\sigma}^2_u,  {\sigma}^2_v, \mathbf{u}_i$m and $\mathbf{v}_k$ are given in Appendix \ref{ap:fullcond}.

For binary affiliation network cases, we can use the same algorithm as above with $\sigma_{\gamma}^2$ set to a fixed value since over-dispersion is not appropriate for generalized linear models for binary responses.  We use this algorithm to fit the model to a binary affiliation network dataset in Section \ref{se:application} with $\sigma_{\gamma}^2 = 1$. 

In addition to the parameters, the dimension of the latent bilinear effects, $t$, is unknown.  Choice of $t$ will generally depend on the goal of the analysis.  If we want to visualize the bilinear terms in order to understand latent structure in an affiliation network, we can simply choose $t=1,\,2$, or $3$.  If the goal is prediction,  we can examine higher dimensions and compare models using the Deviance Information Criterion (DIC; \citealp{Spiegelhalter_etal_2002}), or perhaps formally include $t$ in the model space and employ a reversible jump MCMC algorithm for model fitting \citep{green_1995}.  In Section \ref{se:application}, we try different value for $t$ and report the corresponding DIC.  
Lastly, if we are interested in whether the model captures particular features of the observed network \citep{WF} or in examining particular aspects of lack-of-fit, we can evaluate the model with posterior predictive checks \citep{Besag}.

\section{Application:  Racial Segregation in Extracurricular Activities}
\label{se:application}

\subsection{Motivation and Data Description}\label{se:description}
The presence of members of different racial and ethnic groups within a social unit is referred as \emph{interracial contact}.
Interracial contact in the educational setting is an important social issue, but extant research primarily focuses on the racial composition of the school as a whole or within a classroom.  In high school, however, extracurricular activities play a significant role in students' school experiences, but contact patterns between races within extracurricular activities have been largely unexplored \citep{Mark, charles02}.  Schools that are integrated compositionally may not necessarily result in integrated social interactions if students' social networks are segregated by race/ethnicity.  \citet{mood:2001} finds that schools in which extracurricular activities are integrated by race/ethnicity exhibit lower levels of race/ethnic segregation in friendship networks, suggesting that extracurricular activities play an important role in diversifying the social experiences of youth.  Consequently, accurate characterization of segregation patterns in extracurricular activities by race/ethnicity is necessary in order to understand the features of school social structure that shape actual social interactions and friendship formation.  In this section, we examine interracial contact in high school extracurricular organizations by applying our proposed bilinear mixed effects model to student extracurricular activity network data. 

We consider a binary affiliation network of student participation in extracurricular activities collected by Daniel McFarland as part of his doctoral dissertation at the University of Chicago \citep{data}.  The data is available in \texttt{NetData} R package  \citep{netdata}.   The extracurricular activity data were collected as part of a larger observational study of two high schools that included classroom observations, surveys, school records, and interviews.
We use the data from ``Magnet" High, an elite magnet school located in an inner-city neighborhood of a large Midwestern metropolitan area, whose actual name is redacted to protect the confidentiality of the study participants.  It is an
integrated high school with high-ability students from predominantly lower-income households.  While heterogeneous in racial background, Magnet High is rather homogeneous in terms of student ability.  The extracurricular activity data was collected from information on voluntary participation in clubs and sports in yearbooks.  Gender and racial background on students was ascertained based on yearbook photos, coupled with observation and school records.

\begin{figure}[htp]\begin{center}
\includegraphics[width=6.5in]{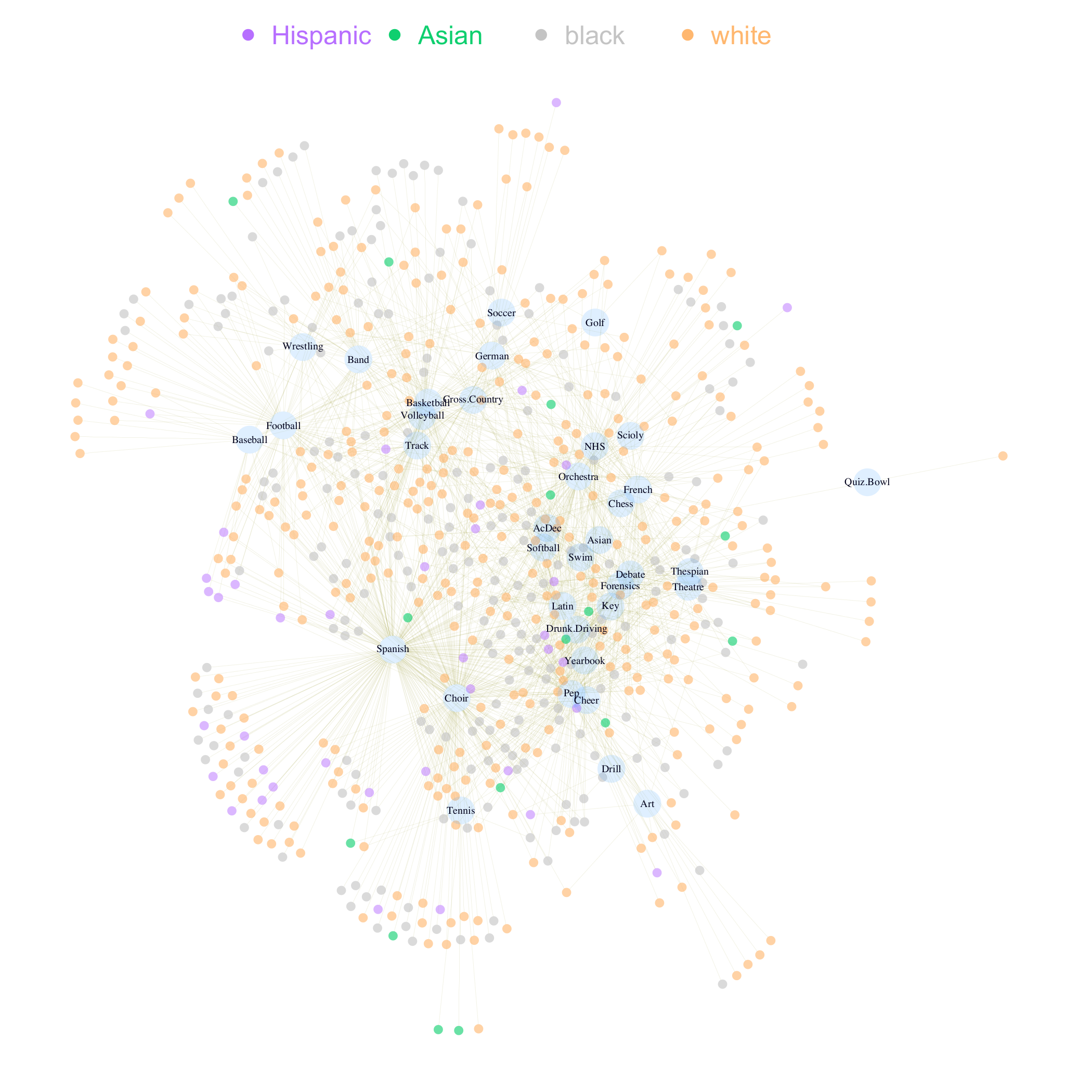}
 \caption{Illustration of the two-mode network of extracurricular activities, with isolated students omitted. The large blue circles represent the $n^e = 37$ activities (events), and the small circles represent the students (actors), where the colors of the plotting symbol indicate the students' races.  If a student participates in an activity, a line is drawn connecting the student and activity. This figure was constructed using functions in the \texttt{iGraph} R package \citep{igraph}.}    
 \label{fig:allplotdata}   
 \end{center}\end{figure}

The full affiliation network data for Magnet High consists of 1295 students and 91 student organizations, in which participation is recorded over three years (1996-1998) along with individual-level attributes of grade, gender, and race. We combine similar clubs together (see Appendix \ref{ap:dataprocessing}) and the newly constructed network has $n^e=37$ activities (events).  We focus on the $n^a=905$ students (actors) in grade 8-12 with non-missing race information listed as Hispanic, Asian, black, and white. We only consider network as it exists in 1996 for our analysis.   The data used in our analysis is shown in Figure \ref{fig:allplotdata}.  While this data is nearly 20 years old, we are not aware of more recent data of a similar nature nor has this data been studied from the perspective of interracial contact.  
 
Magnet High is composed of 6 percent Hispanic, 2 percent Asian, 35 percent black, and 57 percent white  students. Approximately 72 percent of the students participated in at least one activity. A descriptive plot of activity by race is given in Figure Appendix \ref{fig:barplot}, and a summary of the data by race is provided in Table \ref{ta:data}.
\begin{table}
\small
\begin{center}
\begin{tabular}{c | c c c c c c c c c}
\hline
             & Hispanic & Asian & black & white  & All\\
\hline
Number of students  & 54&19&314& 518 & 905 \\
Number of extracurricular activities& 28&17&30&35 & 36\\
Percent Male  & 53.7 & 47.4 &  35.0&  44.8 & 42.0 \\
Percent Female & 46.3 & 52.6 & 65.0 & 55.2 &58.0\\
Percent of participating students & 84.2 & 70.4 &70.7 & 73.0 &  72.3\\
\hline
\end{tabular}
\caption{Various summary statistics by race.  The second row is the number of extracurricular with members of each race.  The last row is the percent of students who participate in at least one extracurricular activity by race.} \label{ta:data}
\end{center}
\end{table}

We fit our proposed bilinear mixed-effects models to the extracurricular affiliation network dataset collected from Magnet High.  There are no dyad-specific covariates in our analysis, so $\beta_d$ corresponds to the intercept.  We use gender and race as the actor-specific covariates.   Taking ``boy" and ``white" as the respective base categories for these two categorical covariates, we have actor effect parameters $\beta^G_a$ (``girl" effect), $\beta_a^B$ (``black" effect), $\beta_a^A$ (``Asian" effect), and $\beta_a^H$ (``Hispanic" effect).  Note that these effects are interpreted relative to the baseline log odds of an activity tie for white boys.  The size of the clubs is the event-specific covariate, with corresponding effect parameter $\beta_e$. 
From our inferences about the positioning of students in the latent social space, we examine the degree of interracial contact in high school extracurricular clubs using techniques from point pattern analysis.

\subsection{Evidence of Higher-Order Dependence}
Before we fit the bilinear mixed-effect model, we check the balance in the data and examine the dependence patterns  as described in Section \ref{se:dp}.  
For this affiliation dataset, the fraction of ties between actors and events is $p_0=0.0705$ (distribution is highly right skewed) and the proportion of balanced cycles, $p$, is $0.8842$.  Under an assumption of independence of the actor/event ties, the expected proportion of positive residual cycles is $p_0^4+(1-p_0)^4+6p_0^2(1-p_0)^2=0.7724$.  If we randomly generate an ${n^a \times n^e}$ matrix with 7.05 percent positive values (+1) and 92.95 percent negative values (-1) 100 times, the greatest observed proportion of balanced cycles in the 100 matrices is $0.8697$. Therefore, we conclude that the observed proportion of balanced cycles is significantly greater than expected under independence (p-value $\approx 0$). 

\subsection{Priors}
Prior distributions for the random effect variances ($ {\sigma}_a^2$, ${\sigma}_e^2$,  ${\sigma}_u^2$, and ${\sigma}_v^2$) are taken to be independent and distributed as $\text{IG}(1,1)$, where $\text{IG}(a,b)$ denotes the inverse gamma distribution with  shape $a>0$ and scale $b>0$.  The priors for $\pmb{\beta}$ are normally distributed $\pmb{\beta} \sim \text{MVN}( \mathbf{0},  \mathbf{I}_{r\times r}).$  The variance of the prior distribution of $\pmb{\beta}$ is small since we are in the logistic regression setting.  

\subsection{Results}
\label{se:results}

The MCMC algorithm described in Section \ref{se:parameter} was run for 150,000 iterations for values of $t=0\, \text{(no bilinear term)},1, \,2, \,3, \, 4, \,5, \,6, \text{and} \,7$.  Trace plots suggest that the Markov chain reaches its stationary distribution well before 100,000 iterations, so we conservatively base our inferences on the last 50,000 iterations. 
DIC \citep{dic}  and alternative DIC \citep{dicalt} are used to assess our models. The results corresponding to different values of $t$ are listed in Table \ref{ta:DIC}.
\begin{table} 
\begin{center}
\begin{tabular}{c | c c c c c c c c c c}
$t$     & $0$ & $1$ & $2$ & $3$  & $4$  & $5$  & $6$ & $7$ \\
\hline
Likelihood        &  -5625 & -5103 & -4893       & -4668 &    -4364 &  -4258   & -4051 & -3950 \\
DIC                      & 9985 & 9533&  9485       &  9516 &    9437 &    9370      & 9287 &9255\\
$\text{DIC}_{alt}$ & 24841 & 25858& 31395&  30809 & 59157 &   76318   &147347 & 208044
\end{tabular}
\caption{DIC  values for models with varying dimension of the components of the bilinear term.} \label{ta:DIC}
\end{center}
\end{table}
In terms of the DIC criterion, model fit generally improves as the dimension of the bilinear terms' components increases, and the largest decrease in DIC occurs when $t$ changes from 0 to 1.  However, the $\text{DIC}_{alt}$ with half the variance of the deviance as an estimate of the number of free parameters in the model keeps increasing and jumps considerably when $t$ increases from $3$ to $4$.   Based on the DIC and $\text{DIC}_{alt}$  and considering our ability to plot in two dimensions, we choose to report inferences on the $t=2$ model.

Table \ref{ta:results} provides the posterior mean and standard deviations for all scalar model parameters when $t=2$.   The 95 percent credible intervals of all actor-specific covariate coefficients cover 0, which implies that student extracurricular participation generally does not appear to depend on gender and race.  As expected, the relationship between club size and the expected log odds of participation is positive with $E[\pmb{\beta}_e \vert \mathbf{Y} ]=0.01$, implying that for every additional member, the odds of a particular individual being in the club increases by one percent (since $e^{0.01} = 1.01$).  
\begin{table} 
\begin{center}
\begin{tabular}{c | c c c c c c c c c c c}
Parameters     & ${\beta}_{d}$ & ${\beta}_a^{G}$ & ${\beta}_a^{b}$ & ${\beta}_a^{A}$ & ${\beta}_a^{H}$ & ${\beta}_e$ & $ {\sigma}_a$ & $ {\sigma}_e$ & $ {\sigma}_u$ & $ {\sigma}_v$ \\
\hline
Mean   &  -4.33  &0.11 & -0.13 & -0.09 & 0.07 & 0.01 & 0.02 & 0.06  &  0.62 & 0.90   \\
SD       &  0.29  & 0.08 & 0.08  & 0.16 & 0.24  &  0.003   & 0.001& 0.01 & 0.34  & 0.41  \\
lower 95\% CI  & -4.91 &-0.05 & -0.29 & -0.40 &-0.39 & 0.01 &0.01 & 0.03 & -0.04 & 0.09  \\
upper 95\% CI & -3.75 & 0.28 & 0.02 & 0.23 & 0.53 & 0.02 & 0.02 & 0.09 & 1.29 & 1.72 \\
\end{tabular}
\caption{Posterior means and standard deviations of model parameters when $t=2$.}\label{ta:results}
\end{center}
\end{table} 

\begin{figure}[htp]\begin{center}
\includegraphics[width=6in]{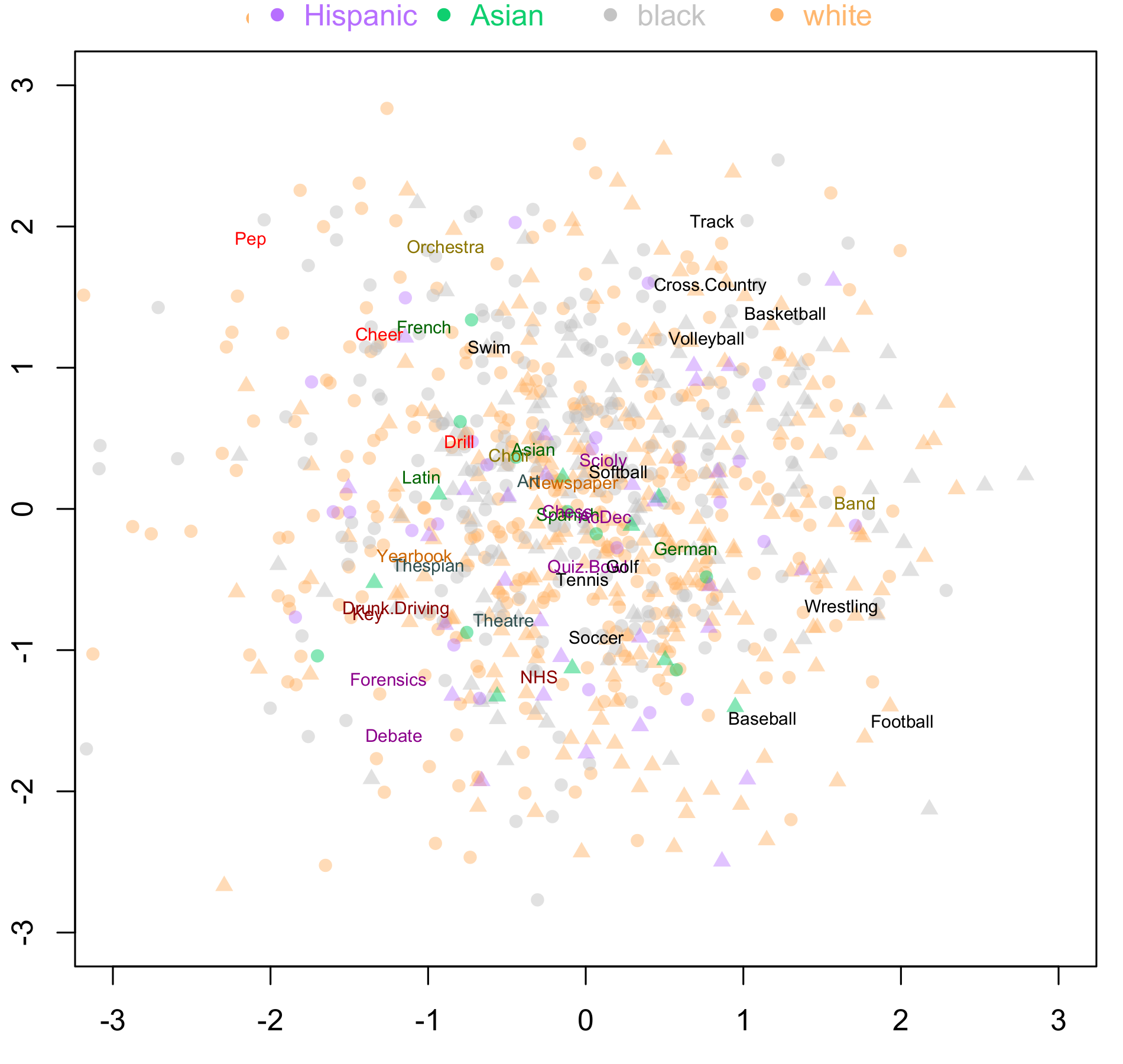}
\caption{Plot of posterior mean of the bilinear terms.  The $\mathbf{v}$s are shown as the locations of the extracurricular activity names with colors corresponding to the categorization in Appendix \ref{ap:dataprocessing}, and the $\mathbf{u}$s corresponding to male and female students are represented as triangles and dots, respectively.}
\label{fig:latentplot}
\end{center}\end{figure}

After eliminating the effects of gender, race, and club size on log odds of a student participating in an activity, we explore the structure of the latent social space through the bilinear term, $\mathbf{u}\mathbf{v}^{\prime}$, which captures dependence between the students through common extracurricular activity profiles.  First, note that the dimension of the bilinear term $\mathbf{u} \mathbf{v}^{\prime}$ is $n^a \times n^e$ and that the bilinear model depends on $\mathbf{u}$ and $\mathbf{v}$ only through the inner products $\mathbf{u}\mathbf{v}^{\prime}$, which is invariant under rotations and reflections of $\mathbf{u}$ and $\mathbf{v}$. To appropriately compare posterior samples of $\mathbf{u}$ and $\mathbf{v}$, we first rotate them to a common orientation using a ``Procrustean" transformation \citep{Sibson} to a Monte Carlo estimate of posterior mean (sample average of the posterior samples), then summarize our inferences by the plot of the posterior mean of $\mathbf{u}$ and $\mathbf{v}$ after rotation which represents the positions of students and activities in the latent social space shown as shown in Figure \ref{fig:latentplot}. 

The structure of the latent social space can be investigated by examining the position of the activity latent vectors, $\mathbf{v}_k$, within this space. To facilitate interpretation in Figure \ref{fig:latentplot}, the extracurricular activities are colored based on the assigned categories of activities listed in Appendix \ref{ap:dataprocessing}.  (Note that these categories were not used in the model fitting.)  The positions of students are shown as points in the latent social space. Triangles represent male students, and dots represent female students.  The plotting symbol color for the students corresponds race, where white, black, Asian and Hispanic students are colored red, black, yellow, and blue,  respectively. 

From Figure \ref{fig:latentplot}, we can see that generally the activities corresponding in the same category tend to be located nearby  each other.  For example, we see that Drill, Cheer, and Pep, the three activities in the Cheer category, are located in the upper left capturing the apparent tendency of students to either participate in all or none of these activities.   On the other hand, there are examples where the clubs do not cluster based on category.  For example, in Music category, Orchestra and Choir locate in the opposite direction from Band indicating a lack of overlap in participants in these groups.  From the distribution of the triangles and dots, we see that males dominate the bottom right quadrant of the plot, which makes intuitive sense since the boys-only sports (Football, Baseball, and Wresting) are oriented in this direction.  We can also see that girls tend to be more active in the Service, News, and Cheer categories.

\subsection{Racial Segregation} \label{se: rs}
To detect patterns of racial segregation within the latent social space,  we use techniques from spatial point pattern analysis.  Here we consider the students as points in a compact subspace, $D$, defined to be the $6 \times 6$ square centered at $(0, 0)$ in $\mathbb{R}^2$.  Each point in $D$ has ``mark" defined by the corresponding student's race.  In order to examine racial segregation of students in terms of the extracurricular activity profiles, after controlling for the relative propensity to participate in activities by race and gender, we look for evidence of clustering by race in $D$.  

\begin{figure}[htp]\begin{center}
\includegraphics[width=7in]{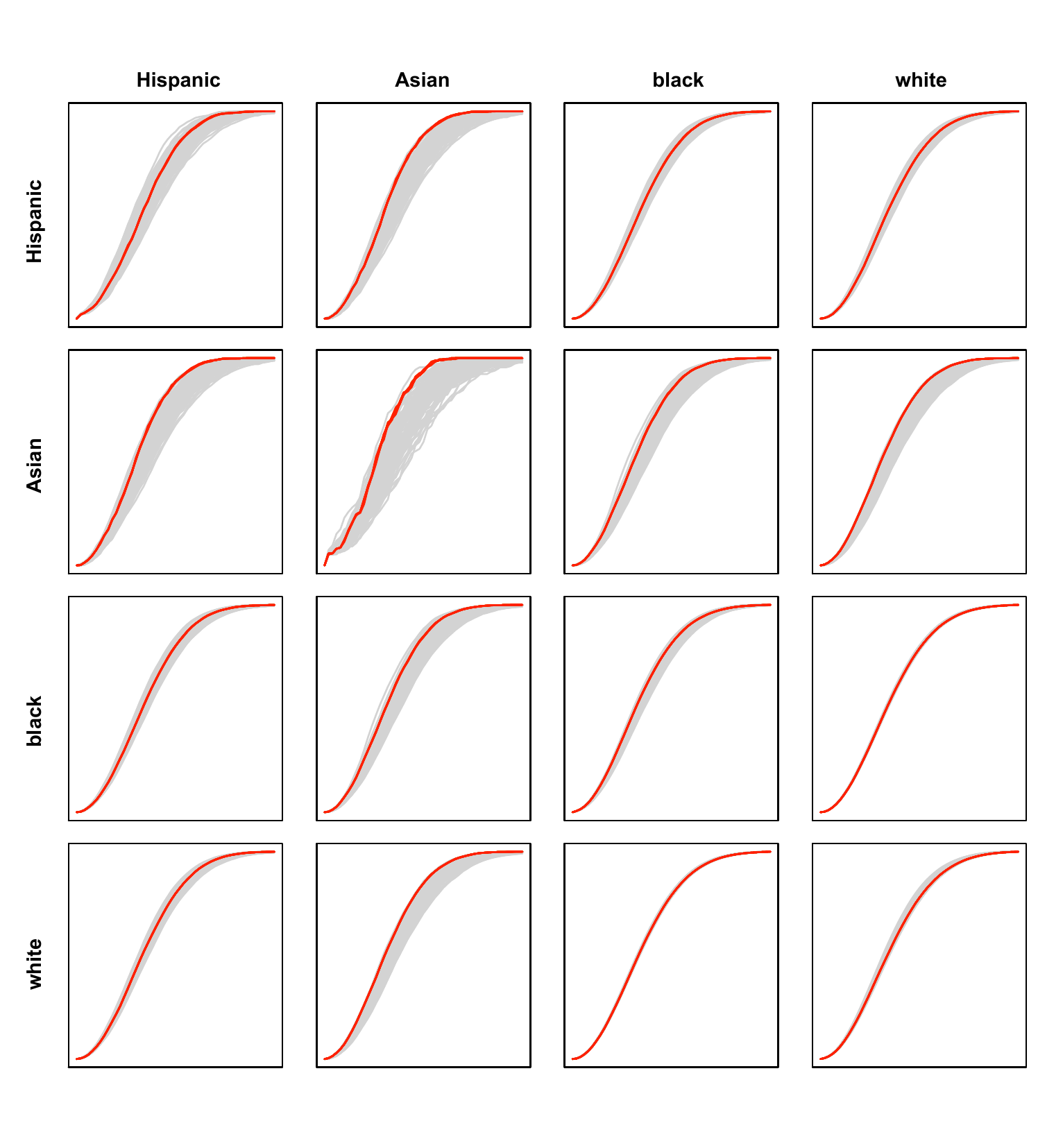}
\caption{Posterior samples of the multi-type $K$-functions for pairs of races (red).  References curves (grey) are computed under randomization of race (see Section \ref{se:results}).}
\label{fig:kfunction}
\end{center}\end{figure}

Our investigation of clustering by race is based on Ripley's multi-type $K$-function (\citealp{diggle}, pages 123-124), where $K_{r_1r_2}(h)$ is defined as the expected number of students of race $r_2$ within a distance $h$ of a typical student of race $r_1$, divided by $\lambda_{r_1r_2}$, where $\lambda_{r_1r_2}=\lambda_{r_1}+\lambda_{r_2}$ is sum of the integrated intensity functions of points with marks $r_1$ and $r_2$ over $D$.
For a particular posterior realization, $m$, let $\mathbf{u}^{[m]}_i$ denote the position of student $i$ in the latent activity space after rotation.  We can estimate $K_{r_1r_2}(h)$ for this realization as
\begin{align*}
\hat{K}_{r_1r_2}^{[m]}(h) &=\frac{1}{\hat{\lambda}^{[m]}_{r_1r_2}}\times \frac{\sum_{i=1}^{n_a}  \sum_{j=1}^{n_a} I(r(i) = r_1\, \& \,r(j) = r_2) I(\vert\mathbf{u}^{[m]}_i - \mathbf{u}^{[m]}_j \vert < h) }{n_{r_1}+n_{r_2}} \\
&=\frac{ \mbox{area}(A)}{(n_{r_1}+n_{r_2})^2}\times \sum_{i=1}^{n^a}  \sum_{j=1}^{n^a} I(r(i) = r_1 \,\& \,r(j) = r_2) I(\vert\mathbf{u}^{[m]}_i - \mathbf{u}^{[m]}_j \vert < h) ,
\end{align*}
where $I(\cdot)$ is the indicator function, $r(\cdot)$ returns the race of the student indexed by the function's argument, $n_{r_1}$ and $n_{r_2}$ are the number of students of race $r_1$ and $r_2$, respectively.  We can estimate $E[K_{r_1r_2}(h) \vert \mathbf{Y}]$ by averaging over the $\hat{K}_{r_1r_2}^{[m]}(h)$ point-wise over $h$.  Alternatively, to get a sense of posterior variability, Figure \ref{fig:kfunction} shows ten posterior multi-type $K$-functions (in red) for each pair of races.  (Note that the differences between these red curves is difficult to discern due to the scale of the plots.)  For reference, we also recomputed the estimated $\hat{K}^{[m]}_{r_1r_2}(h)$ under randomization (sampling without replacement) of the race of students with race either $r_1$ and $r_2$, ten times for each $m$.  These estimated multi-type $K$-functions are shown in grey in Figure \ref{fig:kfunction}.  For all pairs of races, each of the posterior estimated $K$-functions fall within the point-wise as a function of $h$ lower and upper bounds determined by the estimated $K$ functions simulated under randomization, although the Hispanic curves are near the upper bounds.  Therefore, we do not find there to be strong evidence of clustering by race within the latent social space determined by extracurricular activity participation.  Returning to Figure \ref{fig:allplotdata}, we can see that in fact the Hispanic students do not appear to be distributed uniformly in this bipartite graphical summary of the data.  Our work provides a model-based methodology for formally exploring whether or not students are segregated by race in their extracurricular activity profiles.  Lastly, we note that these conclusions are not affected by changing $t$.

\section{Discussion}
\label{se:discussion}
Extracurricular activities play an important role in shaping the social experiences of high school students.  Our contribution is a statistical model that can allow a deeper exploration into the patterns of racial and ethnic segregation in extracurricular activity participation within a school.  Using a regression framework, we are able to control for differences in participation generally by both activity characteristics and attributes of the students.  Our approach allows us to move beyond the examination of specific activities separately.  Through inferences on the bilinear component of our model, we are able to uncover patterns in shared participation across multiple activities.  Thus, unlike previous work which focuses on differences in the amount of participation by race \citep{charles02}, our methodology can uncover differences in the overall patterning of participation by race.  

Our proposed model is based on a generalized linear mixed-effects model with the addition of an inner product of two latent vectors.   As we have shown, this latent structure allows us to capture the types of dependence patterns (balance) usually seen in affiliation networks. Our model improves on existing \textit{ad hoc} methods for analyzing affiliation networks in that we coherently capture the uncertainty in our inferences on the latent structure of a network.  Visualizing this uncertainty is a challenging task, and we plan to work on this important issue in future work.

As is often the case for fitting complex Bayesian models using MCMC, computation can be challenging when the sample size (number of actors and events) is large.  Accordingly, future work will seek to exploit computational tricks to take advantage of sparsity in affiliation networks, as well as explore approximate computation techniques.  

Lastly, the latent structure of our model provides a natural mechanism for incorporating grouping structure in either the actors or events.  For example, in the extracurricular activity example, we could have built in the expected similarity between activities in the same category by placing a hierarchical prior on the activity random effects and the components of the $\mathbf{v}$s.  In this way, we would be able to explain the dependences in affiliation networks both within and across different groups of events.
 
\section*{Acknowledgements}
Support for this work was provided by grants from the National Science Foundation (NSF DMS-1209161), the National Institutes of Health (NIH R01DA032371), the William T. Grant Foundation, and The Ohio State University Institute for Population Research (NIH P2CHD058484).
\bibliography{BilinearModel}		

\newpage

\section*{APPENDICES} 
\appendix
\begin{appendix}
 \renewcommand\thefigure{\thesection.\arabic{figure}}   
\section{Full Conditional Distributions}\label{ap:fullcond}

\subsection*{Full conditional distribution of $(\pmb{\beta}_d, \pmb{\mu}^a,\pmb{\mu}^e)$}
  The full conditional distribution of $(\pmb{\beta}_d, \pmb{\mu}^a ,\pmb{\mu}^e)$ is proportional to 
  \[
  p(\mathbf{z}| \pmb{\beta}_d, \pmb{\mu}^a,\pmb{\mu}^e, {\sigma}^2_{\gamma} ) \times p(\pmb{\mu}^a ,\pmb{\mu}^e | \pmb{\beta}_a, \pmb{\beta}_e,  \sigma^2_a,  \sigma^2_e ) \times p( \pmb{\beta}_d). 
  \]
  
Assume the prior distribution of $\pmb{\beta}_d$ follows a multivariate normal distribution
\[
\pmb{\beta}_d \sim  \text{MVN} ( \pmb{\mu}_{\pmb{\beta}_d}, \pmb{\Sigma}_{\pmb{\beta}_d} ).
\]
We already know that $
\mu^a_i = \pmb{\beta}_a ^{\prime} \mathbf{x}^a_{i}+a_i 
$
and 
$
\mu^e_k = \pmb{\beta}_e ^{\prime} \mathbf{x}^e_{k}+e_k 
$. So, we have
\[
\pmb{\mu}^a,\pmb{\mu}^e \vert \pmb{\beta}_a, \pmb{\beta}_e,  {\sigma}^2_a,  {\sigma}^2_e \sim \text{MVN}( \mathbf{X}_{ae} \pmb{\beta}_{ae} , \pmb{\Sigma}_{ae} ),
\]
where 
$\mathbf{X}_{ae}$ is a $(n^a+n^e)\times2$ matrix, $\pmb{\beta}_{ae}=(\pmb{\beta}_a,\pmb{\beta}_e)^{\prime}$, and $\pmb{\Sigma}_{ae}=\left(  \begin{array}{cc} 
   {\sigma}^2_a \mathbf{I}_{n^a} & 0  \\
  0 &  {\sigma}^2_e \mathbf{I}_{n^e} 
  \end{array}
   \right)$.

Since $\pmb{\beta}_d, \pmb{\mu}^a, \pmb{\mu}^e$ are independent and Gaussian, we can rewrite their joint distribution as
\begin{equation*}
\pmb{\beta}_d, \pmb{\mu}^a , \pmb{\mu}^e \vert \pmb{\beta}_a, \pmb{\beta}_e,  {\sigma}^2_a,  {\sigma}^2_e 
\sim \text{MVN} \left[ 
\left(
\begin{array}{c}
\pmb{\mu}_{\pmb{\beta}_d} \\
\mathbf{X}_{ae} \pmb{\beta}_{ae} 
\end{array} \right) ,  
\left(  \begin{array}{cc} 
  \pmb{\Sigma}_{\pmb{\beta}_d} & 0\\
  0 & \pmb{\Sigma}_{ae}
  \end{array} \right) \right].
\end{equation*}
Let $z_{ik}=\theta_{ik} -\mathbf{u}_i ^{\prime} \mathbf{v}_k = \pmb{\beta}_d ^{\prime} \mathbf{x}^d_{ik} +   {\mu}^a_i+\mu^e_k  + \gamma_{ik} $ so
\begin{equation*}
\mathbf{z} \vert \pmb{\beta}_d, \pmb{\mu}^a,\pmb{\mu}^e, {\sigma}^2_{\gamma} 
\sim\text{MVN} \left[ 
\mathbf{X}_D
\left( \begin{array}{c}  
\pmb{\beta}_d\\
\pmb{\mu}^a\\
 \pmb{\mu}^e
\end{array} \right),  
{\sigma}_{\gamma}^2 \mathbf{I}_{n_{\gamma}}
   \right]
\end{equation*}
It follows that the full conditional distribution $(\pmb{\beta}_d, \pmb{\mu}^a,\pmb{\mu}^e)$ is multivariate normal with the following mean and covariance:

\begin{equation*}
\mathbf{\pmb{\Sigma}} = \left[  \left(
\begin{array}{cc}
\pmb{\Sigma}^{-1}_{\pmb{\beta}_d} & 0 \\
0 & \pmb{\Sigma}^{-1}_{ae}
\end{array} \right)+ 
\mathbf{X}_D^{\prime} \mathbf{X}_D/  {\sigma}^2_{\gamma}  
\right]^{-1}
\end{equation*}

\begin{equation*}
\pmb{\mu} =\pmb{\Sigma} \left[  \left(
\begin{array}{c}
\pmb{\Sigma}^{-1}_{\pmb{\beta}_d} \pmb{\beta}_d \\
\pmb{\Sigma}^{-1}_{ae} \mathbf{X}_{ae} \pmb{\beta}_{ae}
\end{array} \right)+
\mathbf{X}_D ^{\prime} \mathbf{z}/  {\sigma}^2_{\gamma}  \right]
\end{equation*}

\subsection*{Full conditional distribution of $(\pmb{\beta}_a, \pmb{\beta}_e)$}
The full conditional distribution of $(\pmb{\beta}_a, \pmb{\beta}_e)$ is proportional to 
$
p(\pmb{\mu}^a,\pmb{\mu}^e | \pmb{\beta}_a, \pmb{\beta}_e,  {\sigma}^2_a,  {\sigma}^2_e ) \times p(\pmb{\beta}_a, \pmb{\beta}_e). 
$
Assume the combined regression parameter has a multivariate normal prior:
$
(\pmb{\beta}_a, \pmb{\beta}_e) \sim \text{MVN} ( \pmb{\mu}_{\pmb{\beta}_{ae}}, \pmb{\Sigma}_{\pmb{\beta}_{ae}} )
$.

Therefore, the full conditional is a multivariate normal distribution with the following mean and variance:

\begin{equation*}
\mathbf{\pmb{\Sigma}} =  \left(\mathbf{X}^{\prime}_{ae} \pmb{\Sigma}^{-1}_{ae}  \mathbf{X}_{ae} + \pmb{\Sigma} ^{-1}_{\pmb{\beta}_{ae}} \right) ^{-1}
\end{equation*}

\begin{equation*}
\pmb{\mu} =\pmb{\Sigma} \left[ \mathbf{ X}_{ae}^{\prime} \pmb{\Sigma}^{-1}_{ae}\left(
\begin{array}{c}
\pmb{\mu}^a\\
\pmb{\mu}^e
\end{array} \right)+ \pmb{\Sigma}^{-1}_{\pmb{\beta}_{ae}} \pmb{\mu}_{\pmb{\beta}_{ae}} \right].
\end{equation*}

\subsection*{Full conditional distribution of ${\sigma}^2_a,  {\sigma}^2_e $}
We restrict $ {\pmb{\Sigma}}_a= {\sigma}^2_a \mathbf{I}_{n^a \times n^a} $ 
and 
$ {\pmb{\Sigma}}_e= {\sigma}^2_e \mathbf{I}_{n^e \times n^e} $.  For ${\sigma}_a^2 \sim \text{IG} (\alpha_{a1}, \alpha_{a2})$, and $\ {\sigma}_e^2 \sim \text{IG} (\alpha_{e1}, \alpha_{e2})$, the full conditionals are independent and
\[
 {\sigma}_a^2  \vert \pmb{\mu}^a \sim \text{IG} ( n^a/2+ \alpha_{a1}, \alpha_{a2}+  (\pmb{\mu}^a-\mathbf{X}_a \pmb{\beta}_a)^{\prime}(\pmb{\mu}^a-\mathbf{X}_a \pmb{\beta}_a) /2)
\]
and
\[
 {\sigma}_e^2  \vert\pmb{\mu}^e \sim \text{IG} ( n^e/2+ \alpha_{e1}, \alpha_{e2}+ (\pmb{\mu}^e-\mathbf{X}_e \pmb{\beta}_e)^{\prime}(\pmb{\mu}^e-\mathbf{X}_e \pmb{\beta}_e) /2).
\]

\subsection*{Full conditional distribution of $ \sigma^2_{\gamma} $}
We restrict $\pmb{\Sigma}_{\gamma}=\sigma^2_{\gamma}  \mathbf{I}_{n^{\gamma} \times n^{\gamma} } $.
Using prior distribution of 
 $ \sigma_{\gamma}^2 \sim  \text{IG}( \alpha_{{\gamma}1}, \alpha_{{\gamma}2}) $, 
the full conditional distribution of $\sigma_{\gamma}^2$ is
\[
 {\sigma}_{\gamma}^2  \vert \pmb{\beta}_d, \pmb{\mu}^a, \pmb{\mu}^a,  \mathbf{z} \sim \text{IG} 
 \left( \alpha_{\gamma1}+ n^{\gamma}/2,
\alpha_{\gamma2}+\left[ \mathbf{z} -
\mathbf{X}_D
\left( \begin{array}{c}  
\pmb{\beta}_d\\
\pmb{\mu}^a\\
 \pmb{\mu}^e
\end{array} \right) \right] ^{\prime}
\left[ \mathbf{z} -
\mathbf{X}_D
\left( \begin{array}{c}  
\pmb{\beta}_d\\
\pmb{\mu}^a\\
\pmb{\mu}^e
\end{array} \right) \right] /2\right)
\]

\subsection*{Full conditional distribution of $\mathbf{u}_i$}
Let $\theta_{ik} = \pmb{\beta}_d ^{\prime} \mathbf{x}^d_{ik} +  \mu^a _i +\mu^e_k  + \gamma_{ik}+\mathbf{u}_i^{\prime}\mathbf{v}_k$,
as before, and  
$
\hat{\theta}_{ik} = E(\theta_{ik} \vert \pmb{\beta}_d, \mu^a_i, \mu^e_k, x_{ik}^d  )
= \pmb{\beta}_d^{\prime} \mathbf{x}_{ik}^d+  \mu^a _i +\mu^e_k 
$.
Then 
$
e_{ik}  = \theta_{i,k} -\hat{\theta}_{ik}  
 =\mathbf{v_{k}}^{\prime} \mathbf{u}_i  +  \pmb{\gamma}_i
$,
 
Considering the full conditional of $\mathbf{u}_i$, we have

\begin{equation*}
\underbrace{ \left(
\begin{array}{c} 
e_{i,1}^{u}\\
e_{i,2 }^{u}\\
\dots\\
e_{i,n^e}^{u}
\end{array} \right) }_ {\mathbf{e}_i^{u}}
= \underbrace{ 
\left(
\begin{array}{c} 
\mathbf{v}_1\\
\mathbf{v}_2\\
\dots\\
\mathbf{v}_{n^e}
\end{array} \right) }_{\mathbf{v}}  \mathbf{u}_i
+\underbrace{ \left( \begin{array}{c} 
\gamma_{i,1 } \\
\gamma_{i,2 }  \\
\dots\\
\gamma_{i,n^e }  
\end{array} \right)}_{\pmb{\gamma}_{i}^\mathbf{u} },
\end{equation*}
 so $\pmb{e}_i^{u} \vert  \mathbf{v}, \mathbf{u}_i, {\sigma}_{\gamma}^2 \sim \text{MVN}( \mathbf{v} \mathbf{u}_i, {\sigma}_{\gamma}^2 \mathbf{I}_{n_e} ) $. Therefore, sampling $\mathbf{u}_i$ from its full conditional is equivalent to a Bayesian linear regression problem. Assuming the $\mathbf{u}_i$s are \textit{a priori} independent and 
$
\mathbf{u}_i \sim \text{MVN} (0, \pmb{\Sigma}_u),
$
the full conditional of $\mathbf{u}_i$ is multivariate normal with
\[
\pmb{\Sigma}= (\pmb{\Sigma}_u^{-1}+\mathbf{v}^{\prime}\mathbf{v}/ { {\sigma}_{\gamma}^2})^{-1}
\]
and
\[
\pmb{\mu}=\pmb{\Sigma} \mathbf{v}^{\prime} \mathbf{e}_i^u /  {\sigma}_{\gamma}^2. 
\]

\subsection*{Full conditional distribution of $\mathbf{v}_k$}
Similar to the derivation of the full conditional distribution of $\mathbf{v}_i$,  we have
$
e_{ik}  = \theta_{ik} -\hat{\theta}_{ik}  
 =\mathbf{u}_i^{\prime} \mathbf{v}_k  +  \pmb{\gamma}_i
$.
 
Considering the full conditional of $\mathbf{ v}_k$, we have
\begin{equation*}
\underbrace{ \left(
\begin{array}{c} 
e_{1,k} \\
e_{2,k } \\
\dots\\
e_{n^a,k} 
\end{array} \right) }_ {\mathbf{e}_k^\mathbf{v}}
= \underbrace{ 
\left(
\begin{array}{c} 
\mathbf{u}_1\\
\mathbf{u}_2\\
\dots\\
\mathbf{u}_{n^a}
\end{array} \right) }_{\mathbf{u}}  \mathbf{v}_k
+\underbrace{ \left( \begin{array}{c} 
\gamma_{1,k } \\
\gamma_{2,k }  \\
\dots\\
\gamma_{n^a,k}  
\end{array} \right)}_{\pmb{\gamma}_{k}^\mathbf{v} }
\end{equation*} 
$\pmb{e}_k^{v} \vert  \mathbf{u}, \mathbf{v}_i, {\sigma}_{\gamma}^2 \sim \text{MVN}( \mathbf{u} \mathbf{v}_k, {\sigma}_{\gamma}^2 \mathbf{I}_{n_a} ) $.  Therefore, sample $\mathbf{v}_k$ from its full conditional is also equivalent to a Bayesian linear regression problem. Assuming the $\mathbf{v}_k$s are \textit{a priori} independent and each 
$
\mathbf{v}_k \sim \text{MVN} (0, \pmb{\Sigma}_v),
$.
the full conditional of $\mathbf{v}_k$ is multivariate normal with
\[
\pmb{\Sigma}= (\pmb{\Sigma}_v^{-1}+\mathbf{u}^{\prime}\mathbf{u}/ { {\sigma}_{\gamma}^2})^{-1}
\]
and
\[
\pmb{\mu}=\pmb{\Sigma} \mathbf{u}^{\prime} \mathbf{e}_k^v /  {\sigma}_{\gamma}^2. 
\]

\subsection*{Full conditional distribution of $\sigma_u^2, \sigma_v^2$}
We restrict $\pmb{\Sigma}_u=\sigma^2_u \mathbf{I}_{t \times t} $ and $\pmb{\Sigma}_v=\sigma^2_v \mathbf{I}_{t \times t} $ and let $\sigma_u^2 \sim \text{IG} (\alpha_{u1}, \alpha_{u2})$ and  $\sigma_v^2 \sim \text{IG} (\alpha_{v1}, \alpha_{v2})$.  Then the full conditionals are
\[
\sigma_u^2  \vert \mathbf{u} \sim \text{IG} ( n^at/2+ \alpha_{u1}, \alpha_{u2}+\text{trace}(\mathbf{u}^{\prime}\mathbf{u})/2)
\]
and
\[
\sigma_v^2  \vert \mathbf{v} \sim \text{IG} ( n^et/2+ \alpha_{v1}, \alpha_{v2}+ \text{trace}(\mathbf{v}^{\prime}\mathbf{v})/2).
\]

\newpage

\section{Data Processing}\label{ap:dataprocessing}
In our analysis of \citet{data}'s data for year 1996, we collapsed several extracurricular activity categories as described below.  For example, our club ``Pep" includes both members of ``Pep.Club" and ``Pep.Club.Officers."  In addition, we grouped the activities into one of eight types labeled as 1-8 below.  These groups were not used in fitting our model, but were helpful in interpreting our fitted model. 

\begin{enumerate}
  \item Language
  \begin{enumerate}
    \item \emph{Asian}
    \item \emph{Spanish} includes Hispanic.Club, Spanish.Club, Spanish.Club..high.,Spanish.NHS  
    \item \emph{Latin} 
    \item \emph{French} includes French.Club..low., French.Club..high., French.NHS  
    \item \emph{German} includes German.Club, German.NHS 
  \end{enumerate}
  \item Academic Competition
  \begin{enumerate}
    \item \emph{Debate}
    \item \emph{Forensics} includes Forensics, Forensics..National.Forensics.League.  
   \item \emph{Chess} 
   \item \emph{Science.Olympiad }
   \item \emph{Quiz.Bowl} 
    \item \emph{Academic.Decathalon} 
  \end{enumerate}
   \item News
  \begin{enumerate}
    \item \emph{Newspaper}
    \item \emph{Yearbook} includes Yearbook.Contributors, Yearbook.Staff  
  \end{enumerate}
    \item Cheer
  \begin{enumerate}
    \item \emph{Pep} includes Pep.Club, Pep.Club.Officers 
    \item \emph{Drill}
    \item \emph{Cheer} includes Cheerleaders..8th, Cheerleaders..9th, Cheerleaders..Spirit.Squad, Cheerleaders..JV, Cheerleaders..V  
  \end{enumerate}
   \item Service
  \begin{enumerate}
    \item \emph{National Honor Society}
    \item \emph{Drunk.Driving} includes Drunk.Driving, Drunk.Driving.Officers  
    \item \emph{Key}
  \end{enumerate}
    \item Art/Theater
  \begin{enumerate}
    \item \emph{Art}   
    \item \emph{Theatre}
    \item \emph{Thespian}
  \end{enumerate}
  \item Music
  \begin{enumerate}
    \item \emph{Band} includes Band..8th, Band..Marching..Symphonic., Band..Jazz  
    \item \emph{Orchestra} includes Orchestra..8th, Orchestra..Full.Concert, Orchestra..Symphonic     
    \item \emph{Choir} includes Choir..treble, Choir..concert, Choir..women.s.ensemble, Choir..a.capella,       
Choir..chamber.singers,
Choir..vocal.ensemble..4.women.,
Choir..barbershop.quartet..4.men. 
  \end{enumerate}  
    \item Sports
  \begin{enumerate}
    \item \emph{Football} includes Football..8th,
Football..9th,
Football..V   
\item \emph{Soccer}
\item \emph{Volleyball}  includes Volleyball..8th,
Volleyball..9th,
Volleyball..JV,
Volleyball..V  
\item \emph{Basketball} includes Basketball..boys.8th,
Basketball..boys.9th,
Basketball..boys.JV,
Basketball..boys.V,
Basketball..girls.8th,
Basketball..girls.9th,
Basketball..girls.JV,
Basketball..girls.V
\item \emph{Baseball} includes Baseball..JV..10th., Baseball..V   
\item \emph{Softball} includes Softball..JV..10th., Softball..V   
\item \emph{Cross.Country} includes Cross.Country..boys.8th, Cross.Country..girls.8th                ,
Cross.Country..boys.V,
Cross.Country..girls.V   
\item \emph{Golf}
\item \emph{Swim} includes Swim...Dive.Team..boys,
Swim...Dive.Team..girls    
\item \emph{Tennis} includes Tennis..boys.V, Tennis.girls.V
\item \emph{Track} includes Track..boys.8th,
Track..girls.8th,
Track..boys.V,
Track..girls.V 
\item \emph{Wrestling} includes Wrestling..8th, Wrestling..V 
  \end{enumerate}
\end{enumerate}

\newpage
 \setcounter{figure}{0}   
\section{Club affiliations by race}\label{ap:barplot}
\begin{figure}[htp]\begin{center}
\includegraphics[width=6in]{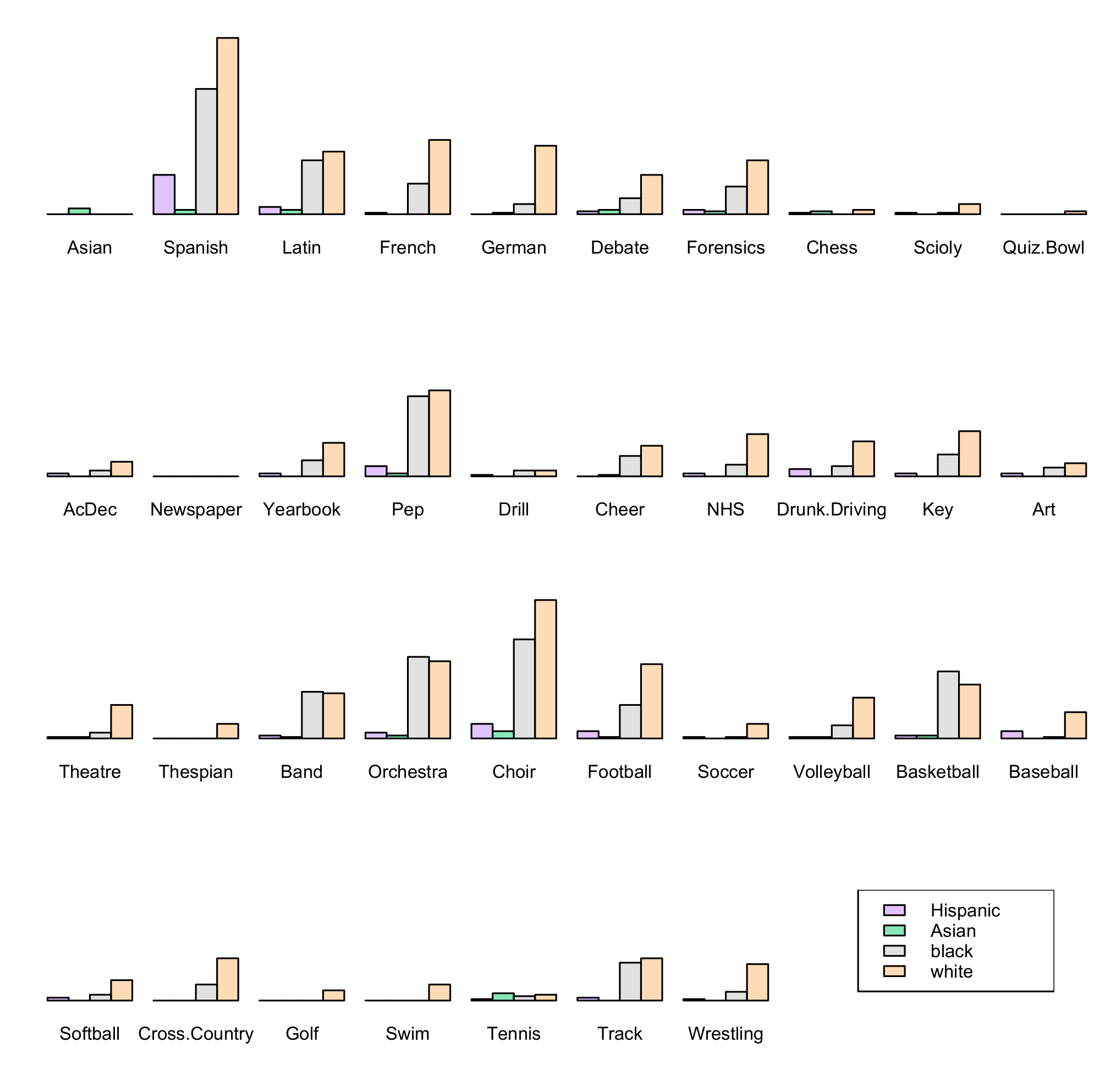}
\caption{Bar plot of the student club affiliations by race.   
}
\label{fig:barplot}
\end{center}\end{figure}

\end{appendix} 
 
\end{document}